# Fast Reversible Learning based on Neurons functioning as Anisotropic Multiplex Hubs

RONI VARDI[1,2,*], AMIR GOLDENTAL[1,*], ANTON SHEININ[3], SHIRA SARDI[1] and IDO KANTER[1,2,+]

[1]Department of Physics, Bar-Ilan University, Ramat-Gan 52900, Israel
[2]Gonda Interdisciplinary Brain Research Center and the Goodman Faculty of Life Sciences, Bar-Ilan University, Ramat-Gan 52900, Israel
[3]Department of Biochemistry and Molecular Biology, Tel Aviv University, Tel Aviv, Israel; Sagol School of Neuroscience, Tel Aviv University, Tel Aviv, Israel

**Abstract** – Neural networks are composed of neurons and synapses, which are responsible for learning in a slow adaptive dynamical process. Here we experimentally show that neurons act like independent anisotropic multiplex hubs, which relay and mute incoming signals following their input directions. Theoretically, the observed information routing enriches the computational capabilities of neurons by allowing, for instance, equalization among different information routes in the network, as well as high-frequency transmission of complex time-dependent signals constructed via several parallel routes. In addition, this kind of hubs adaptively eliminate very noisy neurons from the dynamics of the network, preventing masking of information transmission. The timescales for these features are several seconds at most, as opposed to the imprint of information by the synaptic plasticity, a process which exceeds minutes. Results open the horizon to the understanding of fast and adaptive learning realities in higher cognitive brain's functionalities.

**Introduction.** – Since the pioneering work of Donald Hebb[1], seven decades ago, the common hypothesis in the dynamics of neural networks[2-6] is that the network links, the synapses, are the building blocks of our brain which are responsible for the learning process[7,8]. The basic idea proposed by Hebb was that neurons that fire together, wire together, indicating a local dynamical learning rule to imprint changes in the strengths of the synapses. Following some experimental observations[9], the learning rule was modified and generalized and was found to fluctuate much among different synapses, where a significant change in the strength of the synapses is achieved over time-scales which exceed several minutes. The fast and adaptive learning realities, which is required for high cognitive functions, cannot be attributed to this type of slow learning.

Besides the synaptic plasticity, the plasticity of the neurons[10] was recently found as a source for controlling the network's firing homeostasis and for spontaneous cortical oscillations[11]. The major expressions of the nodal plasticity are the changes in the nodal response timings, the neuronal response latency (NRL), and nodal response probability, resulting in a saturated low-firing frequency, independent of the stimulation frequency. When a neuron is repeatedly stimulated above its characteristic critical frequency (typically ranging between 1 and 30 Hz) its response latency is gradually stretched by several milliseconds until the emergence of a new phase, the intermittent phase, characterized by stochastic response failures which maintain its critical firing frequency[10] (fig. 1b). On the other hand, stimulation of a neuron below its critical frequency leads to consistent response timings and a high response probability. The nodal plasticity varies much among neurons, similar to the synaptic plasticity; however, it is characterized by fast sub-second timescales and by reversibility. These nodal plasticity features were mainly examined using a single stimulation source, while the following scenario of stimulating a neuron from different spatial sources in serial or parallel manner, is the source for the suggested advanced computational paradigm on a network level.

**Results.** –

*Combining Multi-Electrode Array and Patch-Clamp Recordings.* Our experimental results are based on a new available setup, enabling complex extracellular stimulations from a multi-electrode array simultaneously with a patch-clamp recording[12,13] of a single neuron selected from a cultured neural network (fig. 1a and Methods). The dense multi-electrode array enables the stimulation of a neuron from different spatial directions while recording intracellularly its membrane potential and spiking activity (fig. 1).

*Neurons Respond as Multiplex Anisotropic Nodes.* An example of the neuronal response latency (NRL) of a patched neuron, stimulated extracellularly 500 times at 10 Hz by one electrode and then immediately switching to 500 extracellular stimulations at 10 Hz by a different electrode is presented in fig. 2a. The first 500 stimulations demonstrate a typical behavior of a stimulated neuron - the nodal plasticity[10]. The NRL initially increases by several milliseconds, without response failures (light-gray background in figs. 2a,c), until the intermittent phase emerges as a steady phase, characterized by large fluctuations of the NRL around an averaged value accompanied by large fraction of response failures (figs. 2a,c). The switching to stimulations by the second electrode is accompanied by a clear discontinuity in the response characteristics - the NRL of the same neuron stretches again without response failures until the intermittent phase is revisited, as if the first set of stimulations never occurred. Specifically, this realization in fig. 2c indicates that the response probability of another neuron to stimulations from one simulating electrode almost vanishes, while the switching to stimulating by a different electrode results in an increase in the NRL without response failures, until the intermittent phase at a different NRL is achieved and with enhanced neuronal response probability. Results indicate that the NRL and the response probability for a given direction are independent of the very recent inputs from a different direction.

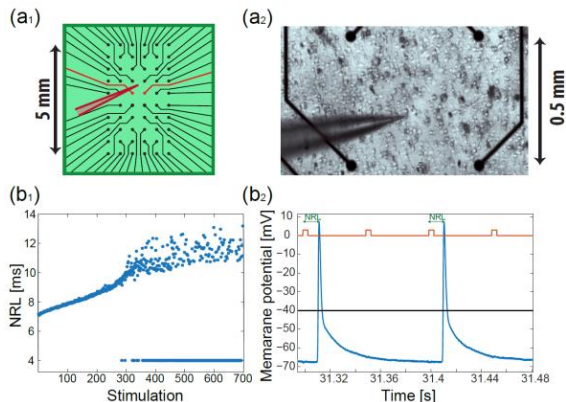

Fig. 1: (Colour on-line) The Apparatus Measurement Combining Multi-Electrode Array and Patch-Clamp Recordings, where the Neuronal Response Latency and the Intermittent Phase are exemplified. ($a_1$) A schema of a 60 multi-electrode array, electrodes are separated by 0.5 mm (or 0.2 mm, see Methods) and with a diameter of 30 micrometers each. The patched recording capillary and two nearby extracellular stimulating electrodes are denoted in red. ($a_2$) An illustration of a limited region of a measured culture, including a capillary of a patched neuron and four extracellular electrodes of the multi-electrode array (see also Methods). ($b_1$) The neuronal response latency (NRL) of a patched neuron. Extracellular stimulations were given at a frequency of 20 Hz, each stimulation has an intensity of 800 mV and a duration of 2 ms. After ~300 stimulations the neuron enters the intermittent phase, where response failures appear (denoted at NRL=4 ms), as well as large fluctuations around an averaged value. ($b_2$) The membrane potential (blue) of a current-clamped neuron at the intermittent phase, demonstrating response failures. Extracellular stimulations are illustrated in orange (scaled). The green arrows at the top show the NRL, the time-lag between a stimulation and its corresponding spike.

Another viewpoint on this discovery is demonstrated by simultaneously stimulating a neuron from two extracellular electrodes, with a frequency of 1 Hz from one and 10 Hz from the other (fig. 2b). Although the arriving stimulations from the two routes mingle at the neuron, the response characteristics of the neuron to each one of them is different and independent. For the 1 Hz stimulations, no response failures occur and the NRL is relatively stable, whereas for the 10 Hz stimulations an increase of several milliseconds in the NRL is observed and a significant fraction of response failures emerge at the intermittent phase, resulting in ~3 Hz firing frequency. This result indicates that the neuron has a different and independent intermittent phase for each stimulating electrode.

An additional aspect of the multiplexing of the neuron is illustrated by *jointly* stimulating through two extracellular electrodes at a frequency of 30 Hz, odd stimulations from the first electrode and even stimulations from the second electrode (fig. 2d). Odd stimulations arriving from the first stimulation electrode almost always generate evoked spikes and result in a firing frequency close to 15 Hz and a moderate increase in the NRL. On the contrary, the even stimulations from the second stimulation electrode result in an NRL increase, and a significant fraction of response failures emerge when the intermittent phase is achieved, leading to a lower firing frequency, below 6 Hz. Results strongly indicate that each stimulation route has a different response probability and a different NRL profile. Note that the results in figs. 2c,d were obtained using cultures in which the synaptic links are inactive due to the addition of a cocktail of synaptic blockers (see Methods) and the spiking activity of the neuron is induced almost exclusively by the extracellular stimulations.

The result that each stimulation route has a different response probability was also examined in the following symmetric stimulation scheduling between the two electrodes and in a "blocked" culture (fig. 2e). A neuron is jointly stimulated by the first stimulation electrode at 1 Hz and by the second one at 5 Hz (fig. $2e_1$), and later the same neuron is jointly stimulated by the first electrode at 5 Hz and by the second electrode at 1 Hz (fig. $2e_2$). In both cases the NRL and the response probability for each electrode are measured, indicating that the maximal firing frequency of the first electrode (blue) is above 5 Hz, whereas the maximal firing frequency of the second electrode (green) is ~1 Hz and with different duration and profile for the NRL. Hence, all features obtained in the case of spontaneously active cultures without the addition of synaptic blockers (figs. 2a,b), were also obtained in the case of "blocked" cultures (figs. 2c-e), indicating that the underlying mechanism is most probable dendritic computation.

*From isotropic Neurons to a Network Composed of Multiplex Anisotropic Neurons.* Results clearly indicate that the response probability of a neuron to stimulations is anisotropic. The source of this phenomenon is unclear; however, it cannot be attributed to the soma itself, which can generate spikes at very high rates without response failures and with negligible changes in the NRL. A neuron is composed of three main elements (fig. 3a) - a single cell body (soma), dendritic trees which roughly speaking are composed of "external hands" that are responsible to collect the incoming signals to the soma, and an axon which transmits the signal from the soma to the synapses of connected neurons (fig. 3b). The origination of this anisotropy has to be attributed, most probably, to the dendrites. In order to achieve a reliable collection of the incoming signals to the neuron, a dendrite has an oriented giant ramified tree-like structure with sub-millimeter diameter and a total length which can reach several millimeters, a hundred times larger than the diameter of the soma (figs. 3a,b). The number of dendritic trees per neuron varies between one, dozens and even more[14-16], organized in specific orientations or semi-isotropic random like structures (fig. 3a). In a traditional picture, the neuron responds to its incoming signal as a scalar, i.e. the neuronal membrane is summing up the incoming signals regardless of their origination, and then relaying it via a threshold unit. Presented results (fig. 2) suggest that the response timing and the response probability depend on the orientation of the incoming signal. In addition, they are unaffected by signals from other directions. *The neuron acts like a hub with multiplex anisotropic features.*

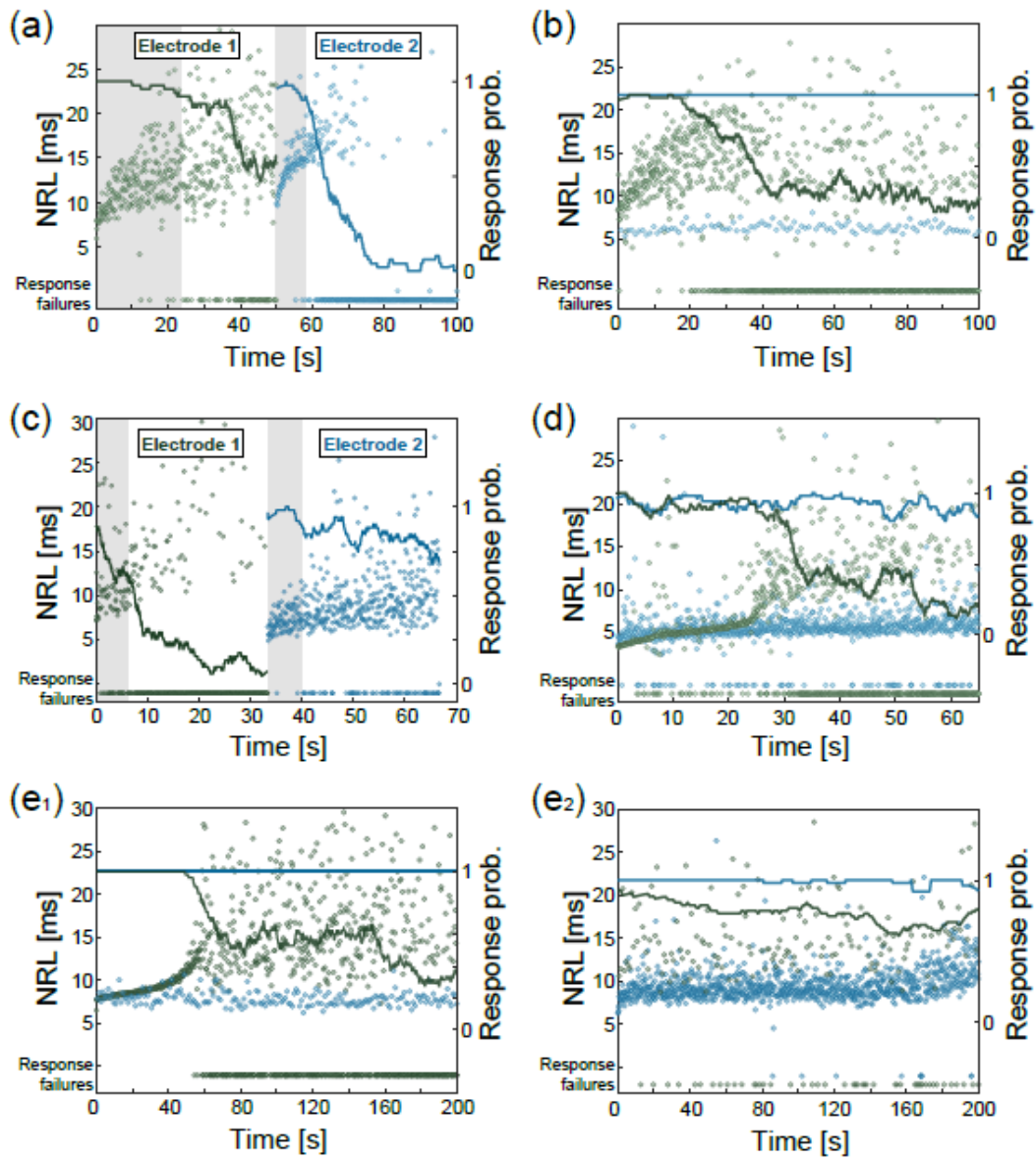

Fig. 2: (Colour on-line) Neurons Respond as Multiplex Anisotropic Nodes. The neuronal response latency (NRL) is denoted by circles, the solid lines indicate the response probability for stimulations and the color code stands for the stimulating electrode. Response failures are denoted as circles at the bottom region of the graphs. (a) An example of the measured response of a patched neuron in an "unblocked" culture, stimulated 500 times at 10 Hz by the first extracellular electrode and then switching, without a delay, to 500 stimulations by the second extracellular electrode (green/blue circles, respectively). The regions of the NRL increase with negligible response failures are denoted by the gray background (see also Methods). (b) The NRL and the response probability for each electrode for a neuron in an "unblocked" culture which is jointly stimulated by the first extracellular stimulation electrode at 1 Hz (blue) and by the second one at 10 Hz (green), with a time offset of 50 ms to avoid simultaneous stimulation by the two electrodes. (c) An example of the measured response of a patched neuron in a "blocked" culture (see Methods), stimulated 500 times at 15 Hz by the first electrode and then switching, without a delay, to 500 stimulations by the second electrode (green/blue circles, respectively). The regions of the NRL increase with negligible response failures are denoted by the gray background. (d) The NRL and the response probability for each electrode for a neuron in a "blocked" culture which is jointly stimulated by two electrodes. The stimulation frequencies of both extracellular electrodes are 15 Hz. As a result, the neuron is overall stimulated, by the two electrodes, at 30 Hz, odd stimulations from the first electrode and even stimulations from the second electrode. ($e_1$) The NRL and the response probability for each electrode for a neuron in a "blocked" culture which is jointly stimulated by the first stimulation electrode at 1 Hz (blue) and by the second one at 5 Hz (green), with a time offset of 100 ms to avoid simultaneous stimulation by the two electrodes. ($e_2$) The same neuron as in ($e_1$), but the first electrode (blue) is now stimulating at 5 Hz and the second electrode (green) at 1 Hz.

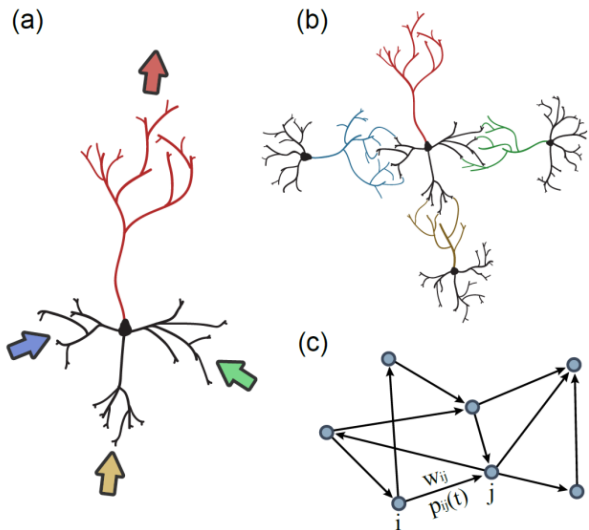

Fig. 3: (Colour on-line) From an Anisotropic Neuron to a Network Composed of Multiplex Anisotropic Nodes. (a) A schema of a pyramidal neuron composed of a soma (black center small object) with three dendritic trees (in black) where each tree has a different input signal denoted by a colored arrow. The output signal travels to other connecting neurons via the axon (red). (b) A schema of three input signals to the central neuron from (a), each signal arrives from a different dendritic tree. The axon of each input signal is colored following the arrows in (a). (c) An artificial network composed of multiplex anisotropic nodes (light blue circles), where each directed link (black arrow) is characterized by its strength, e.g. $w_{ij}$ from node i to node j, and its anisotropic transmission probability, $p_{ij}(t)$, which changes in a much shorter time-scale than w and according to the activity of node i.

*Advanced Functionalities of the Multiplex Anisotropic Node.* The functioning of a neuron as a multiplex anisotropic node[17] raises questions on its possible advanced computational capabilities as well as its impact on cooperative network's features (fig. 3c). On a single neuron level, we present the following three possible advanced functionalities of the multiplex anisotropic node. First, it can assist to maintain the equalization among different information routes, preventing the domination of the nodal activity by one source (fig. 4a). In other types of neurons, where some routes transmit better than others, the anisotropy can lead to a preference of some information routes over the others (fig. 4b). Furthermore, the temporal firing frequency of a neuron can now be much enhanced, since the entire firing pattern is a superposition of all its incoming inputs from the dendritic trees (fig. 4c). The activation of a node in parallel and independently by different information routes enables, with high fidelity, the formation of complex and structured firing patterns, e.g. bursts, (fig. 4c) which are excluded in the scenario of an isotropic node with a single saturated input.

The building block of such a network is a Perceptron (fig. 4d), a multiplex anisotropic node with several independent input routes. Assume several input nodes are very noisy, firing at very high frequencies, consequently inducing, by the overshoot mechanism[18], time slots with vanishing transmission probabilities for those input routes (fig. 4c). Practically, the influence of these inputs on the output signal of this particular node is temporarily excluded, hence the relevant information flows from other input routes are not masked (fig. 4c). This type of a dynamical exclusion has a counterpart in a standard learning process, where the weights connecting the noisy nodes and the output node are weakened and practically vanish. However, this type of link imprinted learning process typically takes many minutes, whereas the presented nodal mechanism for selectively shutting down a particular information route, a link or a bunch of links, occurs on a timescale of seconds. Moreover, after a timescale of several seconds without stimulations, this blocked information route decays from the intermittent phase and the probability for evoked spikes is fully recovered. Hence, this nodal temporal learning process is rapidly emerging and fading out as opposed to the slow irreversible learning process by the links.

**Conclusions.** – The reported neuronal properties call for advanced experiments in order to understand the source of the multiplex anisotropy, in particular, whether its resolution is determined by the entire dendritic tree or by its branches. In addition, it is expected to observe some dependency between the stimulation routes when stimulating from very nearby locations, i.e. adjacent leaves of a dendritic tree, forming some complex multiplexing features, which are important ingredients in modern communication[19].

The characterization of each link by two parameters - the standard link strength[6,20] and the link relaying probability calls for the examination of the mutual interplay between the proposed type of fast reversible information routing and the traditional slow irreversible learning processes[21]. For network science[22-25], and in particular neural networks, the realization of multiplex anisotropic nodes opens the horizons to a better understanding of the brain as well as advanced artificial neural network learning[26] based on the proposed types of networks.

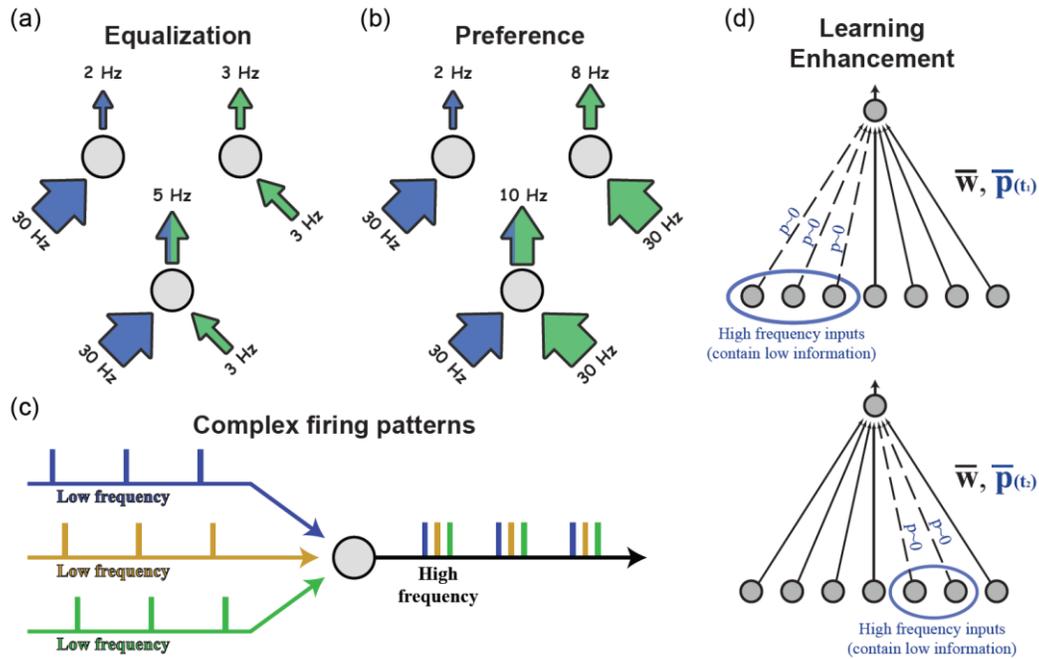

Fig. 4: (Colour on-line) Advanced Functionalities of the Multiplex Anisotropic Node. (a) A schema of a multiplex anisotropic node with two input routes. When stimulated at 30 Hz thorough the left route, the transmission is 2 Hz (top-left) while stimulation at 3 Hz thorough the right route result in a flawless transmission (top-right). The simultaneous stimulation thorough both routes results in an output firing frequency of 5 Hz, with equalization between the two transmitted information routes. (b) Similar to (a) but with a preference to one information route (green) at high stimulation frequencies. (c) Three low-frequency stimulation routes transferred via a neuron with vanishing response failures and forming temporary very high-frequency spiking patterns, which are excluded in the transmission via an isotropic node. (d) A perceptron, a building block of a network, characterized by a set of weights connecting the input units to the output one. A group of input units with temporary very high input frequency (circled in light blue) result in temporary practical vanishing transmission probabilities at $t_1$, $p(t_1) \sim 0$ (upper panel), hence no effect on the output unit, although their weight might be significant. At a different time, $t_2$ (lower panel), the group of input units with vanishing transmission probabilities can be changed fast.

**Methods. –**

*Animals.* All procedures were in accordance with the National Institutes of Health Guide for the Care and Use of Laboratory Animals and Bar-Ilan University Guidelines for the Use and Care of Laboratory Animals in Research and were approved and supervised by the Bar-Ilan University Animal Care and Use Committee.

*Culture preparation.* Cortical neurons were obtained from newborn rats (Sprague-Dawley) within 48 h after birth using mechanical and enzymatic procedures. The cortical tissue was digested enzymatically with 0.05% trypsin solution in phosphate-buffered saline (Dulbecco's PBS) free of calcium and magnesium, and supplemented with 20 mM glucose, at 37∘C. Enzyme treatment was terminated using heat-inactivated horse serum, and cells were then mechanically dissociated. The neurons were plated directly onto substrate-integrated multi-electrode arrays (MEAs) and allowed to develop functionally and structurally mature networks over a time period of 2-4 weeks in vitro, prior to the experiments. The number of plated neurons in a typical network was in the order of 1,300,000, covering an area of about ~5 cm2.

The preparations were bathed in minimal essential medium (MEM-Earle, Earle's Salt Base without L-Glutamine) supplemented with heat-inactivated horse serum (5%), B27 supplement (2%), glutamine (0.5 mM), glucose (20 mM), and gentamicin (10 g/ml), and maintained in an atmosphere of 37∘C, 5% $CO_2$ and 95% air in an incubator.

*Synaptic blockers.* Additional experiments were conducted on cultured cortical neurons that were functionally isolated from their network by a pharmacological block of glutamatergic and GABAergic synapses. For each culture 12 μl of a cocktail of synaptic blockers were used, consisting of 10 μM CNQX (6-cyano-7-nitroquinoxaline-2,3-dione), 80 μM APV (DL-2-amino-5-phosphonovaleric acid) and 5 μM Bicuculline methiodide. This cocktail did not necessarily block completely the spontaneous network activity, but rather made it very sparse. At least half an hour was allowed for stabilization of the effect.

*Stimulation and recording – MEA.* An array of 60 Ti/Au/TiN extracellular electrodes, 30 μm in diameter, and spaced 500 μm or 200 μm from each other (Multi-Channel Systems, Reutlingen, Germany) was used. The insulation layer (silicon nitride) was pre-treated with polyethyleneimine (0.01% in 0.1 M Borate buffer solution). A commercial setup (MEA2100-60-headstage, MEA2100-

interface board, MCS, Reutlingen, Germany) for recording and analyzing data from 60-electrode MEAs was used, with integrated data acquisition from 60 MEA electrodes and 4 additional analog channels, integrated filter amplifier and 3-channel current or voltage stimulus generator. Each channel was sampled at a frequency of 50k samples/s. Mono-phasic square voltage pulses were used, in the range of [−2000, −600] mV and [200, 2000] μs.

*Stimulation and recording – Patch Clamp.* The Electrophysiological recordings were performed in whole cell configuration utilizing a Multiclamp 700B patch clamp amplifier (Molecular Devices, Foster City, CA). The cells were constantly perfused with the slow flow of extracellular solution consisting of (mM): NaCl 140, KCl 3, $CaCl_2$ 2, $MgCl_2$ 1, HEPES 10 (Sigma-Aldrich Corp. Rehovot, Israel), supplemented with 2 mg/ml glucose (Sigma-Aldrich Corp. Rehovot, Israel), pH 7.3, osmolarity adjusted to 300-305 mOsm. The patch pipettes had resistances of 3–5 MOhm after filling with a solution containing (in mM): KCl 135, HEPES 10, glucose 5, MgATP 2, GTP 0.5 (Sigma-Aldrich Corp. Rehovot, Israel), pH 7.3, osmolarity adjusted to 285-290 mOsm. After obtaining the giga-ohm seal, the membrane was ruptured and the cells were subjected to fast current clamp by injecting an appropriate amount of current in order to adjust the membrane potential to about -70 mV. The changes in neuronal membrane potential were acquired through a Digidata 1550 analog/digital converter using pClamp 10 electrophysiological software (Molecular Devices, Foster City, CA). The acquisition started upon receiving the TTL trigger from MEA setup. The signals were filtered at 10 kHz and digitized at 50 kHz.

*Data analysis.* Analyses were performed in a Matlab environment (MathWorks, Natwick, MA, USA). The reported results were confirmed based on at least eight experiments each, using different sets of neurons and several tissue cultures. Evoked spikes were detected by threshold crossing, typically -10 mV, using a detection window of 2.5 - 30 ms following the beginning of an electrical stimulation. The response probability was averaged over windows of 50 stimulations, or the maximal available window at the edges (at least 25).

\*\*\*

We thank Moshe Abeles for stimulating discussions. Invaluable technical assistance by Hana Arnon is acknowledged. This research was supported by the TELEM grant of the Council for Higher Education of Israel.
∗ These authors equally contributed to this work
+ ido.kanter@gmail.com